\documentclass[5p]{elsarticle}
\usepackage{graphicx}
\usepackage{tikz}
\usepackage{mathtools}
\usepackage{amssymb}
\usepackage{chngcntr}

\usepackage{tikz}
\usetikzlibrary{calc}
\usetikzlibrary{backgrounds,arrows,decorations.markings,decorations.pathmorphing,positioning}

\tikzstyle{node} = [draw, circle,fill=blue!20, node distance=3cm,
    minimum height=0em]
\tikzstyle{connection}=[inner sep=0,outer sep=0]    
\tikzset{snake it/.style={decorate, decoration=snake}}

\begin{document}
\title{Polymer simulation by means of tree data-structures and a parsimonious Metropolis algorithm}

\author{Stefan Schnabel}
\ead{stefan.schnabel@itp.uni-leipzig.de}

\author{Wolfhard Janke}
\ead{wolfhard.janke@itp.uni-leipzig.de}

\address{Institut f\"ur Theoretische Physik, Universit\"at Leipzig, Postfach 100920, 04009 Leipzig, Germany}
\date{\today}




\begin{abstract}

We show how a Monte Carlo method for generating self-avoiding walks on lattice geometries which employs a binary-tree data structure can be adapted for hard-sphere polymers with continuous degrees of freedom. Data suggests that the time per Monte Carlo move scales logarithmically with polymer size. We combine the method with a variant of the Metropolis algorithm and preserve this scaling for Lennard-Jones polymers with untruncated monomer-monomer interaction. We further show how the replica-exchange method can be adapted for the same purpose. 

\end{abstract}

\maketitle

\section{Introduction}

The simplest model for a polymer chain realizing nothing but its linear geometry is provided by a random walk, e.g., a random path on a cubic lattice. While these objects can be treated very easily with analytical methods, they do not posses an abundance of realistic features and real polymers behave similar to random walks only at the $\Theta$-point. If, as a step towards more realistic representations, excluded volume interaction is to be included, the simplest model is the self-avoiding walk. It is similar to the random walk, but any lattice site may only be visited once. Different parts of the polymer are not allowed to occupy the same space. Now the problem becomes more difficult and since in three dimension the Flory exponent -- signifying the scaling behavior of geometric quantities such as the end-to-end distance or the radius of gyration -- as well as corrections to scaling are not exactly known, numerical methods are applied.

Ideally, in order to determine expectation values of observables of interest, one would like to sum over all possible walks and algorithms that can efficiently generate these are, therefore, strongly desired. While simple recursive methods on the simple-cubic lattice require days to generate all walks up to 20 steps, much more sophisticated techniques have been applied in order to enumerate walks of 36 steps \cite{Schram}. Monte Carlo (MC) computer simulation methods introduce a statistical uncertainty, but are able to investigate much larger systems and until not too long ago $N\approx10^5$ was the state of the art. Recently, however, Clisby introduced an ingenious technique \cite{Clisby1,Clisby2,Clisby3} for the investigation of self-avoiding walks by means of a binary tree representation simulating walk of length $N\approx 3\times10^7$. One of the technique's main features is a variable resolution; if possible large parts of the walk are treated as single units that are only `zoomed-into' if the situation at hand demands it. The first part of this study is dedicated to adapting this technique to off-lattice hard-sphere polymers.

The effect of solution on a polymer is often modeled implicitly by an attractive term in the Hamiltonian. The transition from a bad solvent with the polymer in a collapsed, globular state towards a good solvent with configurations resembling swollen coils then corresponds to a change from low temperature where the interaction has a strong influence to high temperature where it is rather unimportant. The most commonly used potential is the 12-6 Lennard-Jones potential, where both the repulsive and the attractive component are polynomial with exponents 12 and 6, respectively. Since this potential exerts forces at all distances the recalculation of the energy of the polymer after a modification of the configuration is of complexity $O(N^2)$. In order to achieve a better performance the interaction is often truncated ensuring that beyond a certain distance the forces become zero. We show how this can be avoided by combining the tree-like data structure with an implementation of the Metropolis algorithm that does not require complete information.

In this paper we discuss in section 2 the models under consideration and explore in section 3 the mathematical properties of the transformations that are used to affect conformational updates. In section 4 our version of Clisby's binary trees is described followed by its application to hard-sphere polymers in section 5. We then introduce a variant of the Metropolis algorithm in section 6 which permits the simulation of polymers with untruncated interactions as described in section 7, where we also briefly discuss how the replica exchange method can benefit from the same ideas. We consider polymers with truncated interaction in section 8 and present some conclusions in section 9. Finally, in the appendix we describe a procedure for different types of updates.


\section{Models}
\subsection{Hard-sphere polymer}

We first consider a freely-jointed chain with hard spheres, i.e., the continuum model that most closely resembles the self-avoiding walk on a lattice. 

The positions of monomers $x_1,\dots,x_N$ are given by
\begin{equation}
\mathbf{x}_k=\sum_{i=1}^{k-1}\mathbf{b}_i,
\end{equation}
where the bond vectors have a constant length $|\mathbf{b}_i|=b$ and the distance between any two monomers has a lower bound
\begin{equation}
|\mathbf{x}_k - \mathbf{x}_l |\ge D
\end{equation}
corresponding to the diameter of the hard spheres. All values from $D=0$ (the ideal chain) to $D=b$ are in principle possible. However, for the latter case some caution is required. Since continuum coordinates carry rounding errors, calculated distances between adjacent monomers will deviate slightly from the theoretical value $b$. If the model is to be investigated for $D=b$, this can lead to false overlaps if bonds are smaller than $b$ and consequently to needless rejections of proposed Monte Carlo moves. For the polymer length we considered $(N<10^7)$ the errors of bond length did not exceed $10^{-7}b$ and it is possible to use $D\le(1-10^{-6})b$ ensuring that $|\mathbf b_i|>D$ regardless of numerical errors. Alternatively, a few lines of additional code can exempt  pairs of adjacent monomers from the testing with little effect on overall performance.

\subsection{Lennard-Jones polymer}

For the second model we abandon the hard spheres in favor of a 12-6 Lennard-Jones interaction
\begin{equation}
U_{\rm LJ}(r)=4\epsilon\left[ \left(\frac\sigma r\right)^{12} - \left(\frac\sigma r\right)^{6} \right]
\end{equation}
acting between all pairs of monomers. The Hamiltonian thus reads
\begin{equation}
\mathcal{H}=\sum_{i=1}^{N-1}\sum_{j=i+1}^{N} U_{\rm LJ}( \mathbf{x}_i - \mathbf{x}_j ).
\end{equation}
The performance of the method depends on the choice of $\sigma$, with smaller values leading to faster simulations. For the simulations presented in this paper we chose $\sigma = b/2^{1/6}$ such that the minimum of $U_{\rm LJ}(r)$ coincides with $b$.

Note that for both models we have $N$ monomers connected by $N-1$ bonds as opposed to the notation of $N$ steps and $N+1$ occupied sites that is often used in the context of random walks.


\section{Transformations}

During the simulation we modify the polymer configuration using transformations that we derive as follows. We choose a $k\in\{1,\dots,N-1\}$ and a random axis through $\mathbf{x}_k$ perpendicular to $\mathbf{b}_k=\mathbf{x}_{k+1}-\mathbf{x}_k$. The rotation angle is drawn such that as a result the new position $\mathbf{x}'_{k+1}$ is at a random position on the sphere of radius $|\mathbf{b}|$ around $\mathbf{x}_k$. For the pivot update we apply the rotation that is given by this axis and the angle to all monomers $\mathbf{x}_{k+1},\dots,\mathbf{x}_N$. We also use a bond-rotation update which modifies $\mathbf{x}_{k+1}$ in the same way, but keeps all bonds $\mathbf{b}_i$ with $i\neq k$ unchanged such that $\mathbf{x}'_i|_{i>k+1}=\mathbf{x}_i+\mathbf{x}'_{k+1}-\mathbf{x}_{k+1}$.


In general the rotation of a monomer position $\mathbf{x}$ around an axis that passes through the point $\mathbf{p}$ is given by
\begin{equation}
\mathbf{x}'=\mathcal{R}(\mathbf{x}-\mathbf{p})+\mathbf{p}=\mathcal{R}\mathbf{x}-\mathcal{R}\mathbf{p}+\mathbf{p},
\end{equation}
where $\mathcal{R}$ is the rotation matrix. Hence
\begin{equation}
\mathbf{x}'=\mathcal{R}\mathbf{x}+\mathbf{b},
\label{eqn:trans}
\end{equation}
with
\begin{equation}
\mathbf{b}=\mathbf{p}-\mathcal{R}\mathbf{p}.
\end{equation}
It is easy to see that reflections on arbitrary planes as well as simple shifts by some vector can also be expressed in this form.

Applying two transformations consecutively,
\begin{eqnarray}
\mathbf{x}'' &=& \mathcal{R}_2\mathbf{x'}+\mathbf{b}_2,\\
             &=& \mathcal{R}_2\left(\mathcal{R}_1\mathbf{x}+\mathbf{b}_1 \right)+\mathbf{b}_2,\\
             &=& \left(\mathcal{R}_2\mathcal{R}_1\right)\mathbf{x}+\left(\mathcal{R}_2\mathbf{b}_1+\mathbf{b}_2\right),
\end{eqnarray}
we obtain the same structure as in (\ref{eqn:trans}) which means that it is easily possible to `multiply' two transformations before applying them to the monomer:
\begin{eqnarray}
\mathcal{T}_1\equiv\{\mathcal{R}_1,\mathbf{b}_1\},\quad \mathcal{T}_2\equiv\{\mathcal{R}_2,\mathbf{b}_2\},\label{eqn:trans_def} \\ \mathcal{T}_1\bullet\mathbf{x}\coloneqq\mathcal{R}_1\mathbf{x}+\mathbf{b}_1, \\
\mathcal{T}_2\bullet(\mathcal{T}_1\bullet\mathbf{x})=(\mathcal{T}_2\circ\mathcal{T}_1)\bullet\mathbf{x},
\end{eqnarray}
with
\begin{equation}
\mathcal{T}_2\circ\mathcal{T}_1\coloneqq\{\mathcal{R}_2\mathcal{R}_1,\mathcal{R}_2\mathbf{b}_1+\mathbf{b}_2\}.
\end{equation}
It is also easy to see that this operation is associative
\begin{equation}
(\mathcal{T}_3\circ\mathcal{T}_2)\circ\mathcal{T}_1 = \mathcal{T}_3\circ(\mathcal{T}_2\circ\mathcal{T}_1).
\label{eqn:asso}
\end{equation}


\section{Binary tree}

A few years ago Clisby \cite{Clisby1} has introduced the binary tree as a fundamental data structure for the simulation of self-avoiding walks on lattices and has achieved most impressive results. Strongly inspired by his ground-breaking work we adapt this approach for off-lattice polymers.

\begin{figure}
\begin{center}
\begin{tikzpicture}[level/.style={sibling distance=7em/#1}, every node/.style={transform shape}]
\node [circle,draw,minimum size=1cm, fill=blue!20] (A) {A}
    child { node [circle,draw,minimum size=1cm,fill=blue!20] {B} 
      child { node [circle,draw,inner sep=0pt,minimum size=1cm,fill=red!20] {$k$} 
      }
      child { node [circle,draw,inner sep=0pt,minimum size=1cm,fill=red!20] {$k+1$} 
      }
    }
    child { node [circle,draw,inner sep=0pt,minimum size=1cm,fill=red!20] {$k+2$} 
    };
\path[draw] (0,0.5) -- (0,0.9);

\node [circle,fill=black,inner sep=0pt,minimum size=0.1cm] (X1) at (3,0) {};
\node [right of=X1,node distance =.3cm]  {$\mathbf{x}_k$};

\node [circle,fill=black,inner sep=0pt,minimum size=0.1cm] (X2) at (4,-1) {};
\node [left of=X2,node distance =.4cm]  {$\mathbf{x}_{k+1}$};

\node [circle,fill=black,inner sep=0pt,minimum size=0.1cm] (YB) at (3.5,-0.5) {};
\node [right of=YB,node distance =.3cm]  {$\mathbf{y}_B$};
\node [circle,draw,inner sep=0pt,minimum size=1.5cm] (Y2) at (3.5,-0.5) {};
\node [circle,fill=black,inner sep=0pt,minimum size=0.1cm] at (3,0) {};
\path[draw] (3.5,-0.5) -- node[anchor=north] {$r_B$} (2.75,-0.6);

\node [circle,fill=black,inner sep=0pt,minimum size=0.1cm] (X3) at (4,-3.04) {};
\node [above of=X3,node distance =.3cm]  {$\mathbf{x}_{k+2}$};
\node [circle,draw,inner sep=0pt,minimum size=3.4cm] (Y1) at (3.8,-1.4) {};
\node [circle,fill=black,inner sep=0pt,minimum size=0.1cm] (YA) at (3.8,-1.4) {};
\node [right of=YA,node distance =.3cm]  {$\mathbf{y}_A$};
\path[draw] (3.8,-1.4) -- node[anchor=north] {$r_A$} (2.13,-1.7);

\end{tikzpicture}
\end{center}
\caption{\small{\label{fig:node_sphere_data} Left: A subtree with three monomers represented by the leaf-nodes $k$,$k+1$, and $k+2$ and two internal nodes. Right: The contained geometric information. Monomer positions $\mathbf{x}_{k},\mathbf{x}_{k+1},\mathbf{x}_{k+2}$ and sphere parameters $\mathbf{y}_A,r_A$ and $\mathbf{y}_B,r_B$.}}
\end{figure}
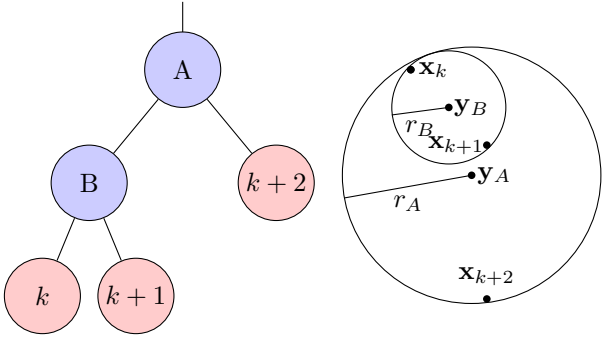

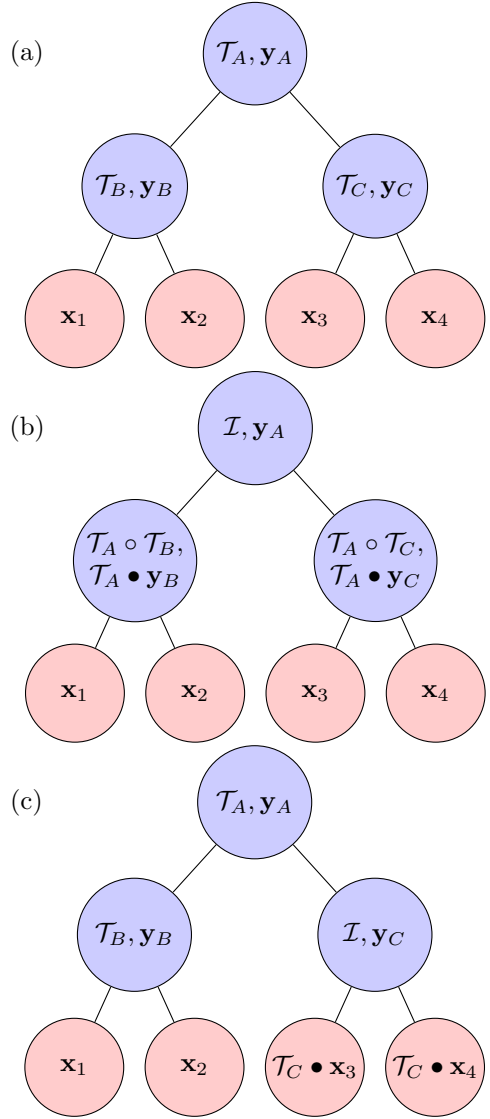
\begin{figure}
\begin{center}
\begin{tikzpicture}[level/.style={sibling distance=9em/#1,level distance=50pt}, every node/.style={transform shape}]
\node [circle,draw,minimum size=1cm, fill=blue!20] (A) {$\mathcal{T}_A,\mathbf{y}_A$}
    child { node [circle,draw,minimum size=1cm,fill=blue!20] {$\mathcal{T}_B,\mathbf{y}_B$} 
      child { node [circle,draw,minimum size=1.3cm,fill=red!20] {$\mathbf{x}_1$} 
      }
      child { node [circle,draw,minimum size=1.3cm,fill=red!20] {$\mathbf{x}_2$} 
      }
    }
    child { node [circle,draw,minimum size=1cm,fill=blue!20] {$\mathcal{T}_C,\mathbf{y}_C$} 
      child { node [circle,draw,minimum size=1.3cm,fill=red!20] {$\mathbf{x}_3$} 
      }
      child { node [circle,draw,minimum size=1.3cm,fill=red!20] {$\mathbf{x}_4$} 
      }
    };
\node [left of=A,node distance=3cm] {(a)};
\end{tikzpicture}

\begin{tikzpicture}[level/.style={sibling distance=9em/#1,level distance=50pt}, every node/.style={transform shape}]
\node [circle,draw,minimum size=1.5cm, fill=blue!20] (B) {$\mathcal{I},\mathbf{y}_A$}
    child { node [circle,draw,minimum size=1cm,inner sep=-3pt,fill=blue!20] {\begin{tabular}{c}$\mathcal{T}_A\circ\mathcal{T}_B,$\\$\mathcal{T}_A\bullet\mathbf{y}_B$\end{tabular}} 
      child { node [circle,draw,minimum size=1.3cm,fill=red!20] {$\mathbf{x}_1$} 
      }
      child { node [circle,draw,minimum size=1.3cm,fill=red!20] {$\mathbf{x}_2$} 
      }
    }
    child { node [circle,draw,minimum size=1cm,inner sep=-3pt,fill=blue!20]  {\begin{tabular}{c}$\mathcal{T}_A\circ\mathcal{T}_C,$\\$\mathcal{T}_A\bullet\mathbf{y}_C$\end{tabular}} 
      child { node [circle,draw,minimum size=1.3cm,fill=red!20] {$\mathbf{x}_3$} 
      }
      child { node [circle,draw,minimum size=1.3cm,fill=red!20] {$\mathbf{x}_4$} 
      }
    };
\node [left of=B,node distance=3cm] {(b)};
\end{tikzpicture}

\begin{tikzpicture}[level/.style={sibling distance=9em/#1,level distance=50pt}, every node/.style={transform shape}]
\node [circle,draw,minimum size=1.5cm, fill=blue!20] (C) {$\mathcal{T}_A,\mathbf{y}_A$}
    child { node [circle,draw,minimum size=1.5cm,fill=blue!20] {$\mathcal{T}_B,\mathbf{y}_B$} 
      child { node [circle,draw,minimum size=1.3cm,fill=red!20] {$\mathbf{x}_1$} 
      }
      child { node [circle,draw,minimum size=1.3cm,fill=red!20] {$\mathbf{x}_2$} 
      }
    }
    child { node [circle,draw,minimum size=1.5cm,fill=blue!20] {$\mathcal{I},\mathbf{y}_C$} 
      child { node [circle,draw,minimum size=1.3cm,inner sep=-3pt,fill=red!20] {$\mathcal{T}_C\bullet\mathbf{x}_3$} 
      }
      child { node [circle,draw,minimum size=1.3cm,inner sep=-3pt,fill=red!20] {$\mathcal{T}_C\bullet\mathbf{x}_4$} 
      }
    };
\node [left of=C,node distance=3cm] {(c)};
\end{tikzpicture}
\end{center}
\caption{\small{\label{fig:tree_4} (a) A binary tree for a polymer with four monomers with transformations in the internal nodes. (b) The tree in (a) with the transformation $\mathcal{T}_A$ pushed down. (c) The tree in (a) with the transformation $\mathcal{T}_C$ pushed down. Here, $\mathcal{I}$ stands for the identity or absence of a transformation.}}
\end{figure}

All data is organized in a binary tree where the leaves represent individual monomers and any internal node, i.e., a node that is not a leaf, provides a coarse-grained representation of its children: Each node contains among other data the parameters for a sphere that comprises all monomers in the sub-tree to which it is root (Fig.~\ref{fig:node_sphere_data}). Such a representation serves two purposes. On the one hand it allows to ensure that distinct parts of the polymer represented by different nodes do not overlap, since a sufficient (although not necessary) condition is that the respective spheres do not intersect.
On the other hand modifications to the polymer can be applied at a level of low resolution to nodes high in the tree. To that end each internal node is able to store a transformation of the shape defined in (\ref{eqn:trans_def}) that applies to all nodes in its sub-tree, i.e., the nodes constituting its collective offspring, except the node itself. Consider for instance the tree in Fig.~\ref{fig:tree_4}(a). Although the monomer position stored in the lower left node is $\mathbf{x}_1$, the actual position of the first monomer is given by $\mathcal{T}_A\bullet(\mathcal{T}_B\bullet\mathbf{x}_1)$, and the position of the center of sphere that belongs to node B is actually $\mathcal{T}_A\bullet\mathbf{y}_B$. Only when a need to access a certain position in a node arises the respective transformations in the `ancestor'-nodes are applied. This is either done separately outside the tree or by pushing down transformations as depicted in Figs.~\ref{fig:tree_4}(b,c).

Let us summarize which data has to be stored in a single node:
\begin{itemize}
\item{links relevant for the geometry of the tree, i.e., links to the parent-node and the two children,}
\item{parameters for a sphere that contains all monomers in the sub-tree to which the node is root,}
\item{the data for a transformation that is to be applied to all nodes in the nodes sub-tree, but not to the node itself,}
\item{additional information, e.g., index of the node or size of the sub-tree.}
\end{itemize}

Since the underlying data structure is a binary tree it is natural although not required to chose system sizes that are powers of two.


\section{Simulating the hard-sphere polymer}

\begin{figure}
\begin{center}
\begin{tikzpicture}[level/.style={sibling distance=10em/#1,level distance=30pt}, every node/.style={transform shape},every node/.style={circle},circle dotted/.style={dash pattern=on .05mm off 4mm, line cap=round}]
\node [minimum size=0.9cm,draw, fill=blue!20] (A) {R}
    child { node [minimum size=0.9cm,draw,fill=blue!20] {l} 
      child { node [minimum size=0.9cm,draw,fill=blue!20] {ll} 
        child { node [minimum size=0.1cm] {} }
        child { node [minimum size=0.1cm] {} }
      }
      child { node [minimum size=0.9cm,draw,fill=blue!20] {lr} 
        child { node [minimum size=0.1cm] {} }
        child { node [minimum size=0.1cm] {} }
      }
    }
    child { node [minimum size=0.9cm,draw,fill=blue!20] {r} 
      child { node [minimum size=0.9cm,draw,fill=blue!20] {rl} 
        child { node [minimum size=0.1cm] {} }
        child { node [minimum size=0.1cm] {} }
      }
      child { node [minimum size=0.9cm,draw,fill=blue!20] {rr} 
        child { node [minimum size=0.1cm] {} }
        child { node [minimum size=0.1cm] {} }
      }
    };
\path[draw,color=gray] (0,-1.) -- (0,-5);
\node [circle,minimum size=0.9cm,draw,fill=red!20] at (-3,-4.5) {$\mathbf{x}_1$};
\path[draw] (-3,-4.05) -- (-3,-3.77);

\node [circle,minimum size=0.9cm,draw,fill=red!20] at (-0.6,-4.5) {$\mathbf{x}_k$};
\path[draw] (-0.6,-4.05) -- (-0.6,-3.75);

\node [circle,minimum size=0.9cm,inner sep=0pt,draw,fill=red!20] at (0.6,-4.5) {$\mathbf{x}_{k+1}$};
\path[draw] (0.6,-4.05) -- (0.6,-3.64);

\node [circle,minimum size=0.9cm,draw,fill=red!20] at (3,-4.5) {$\mathbf{x}_N$};
\path[draw] (3,-4.05) -- (3,-3.62);

\draw[line width=1mm,circle dotted] (-2.4,-4.5) -- (-1.1,-4.5);
\draw[line width=1mm,circle dotted] (2.4,-4.5) -- (1.1,-4.5);

\draw[snake it] (-0.1,-3.0)--(-3.3,-3.0);
\draw[snake it] (-0.1,-3.7)--(-3.3,-3.7);
\draw[snake it] (0.1,-3.0)--(3.3,-3.0);
\draw[snake it] (0.1,-3.7)--(3.3,-3.7);
\end{tikzpicture}
\end{center}
\caption{\small{\label{fig:tree_upd} The binary tree during an update.}}
\end{figure}
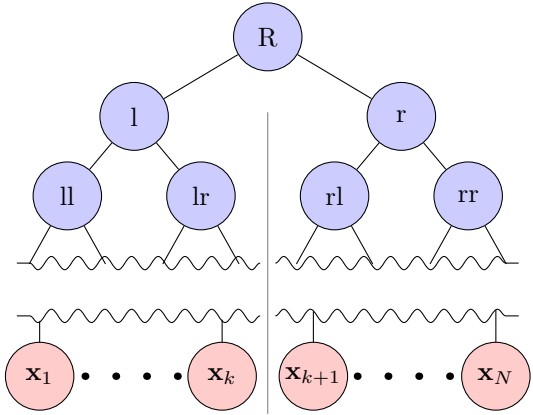

In order to perform an update of the polymer it is convenient to rearrange the tree. Consider Fig.~\ref{fig:tree_rot}, where versions of a section of a tree are displayed. Transitions between them are called tree-rotations. With respect to the parts of the polymer they contain, during these operations only the nodes B and D are altered: Node B has the children A and D while node B' links to A and C, while the children of D are C and E as opposed to B' and E for D'. This means that during such an operation one has to recalculate the spheres in the nodes B' and and D' if going from left-to-right in Fig.~\ref{fig:tree_rot} or in D and B when moving right-to-left. First, however, it is important to take care of potentially stored transformations. The easiest way to do this is to push down the transformations in B and D (B' and D') such that both nodes do not hold a transformation when the actual tree-rotation is performed. Note that the horizontal order of the nodes is not affected. From left to right the nodes read A,B,C,D,E or A,B',C,D',E. Furthermore, each internal node, i.e., each node that does not represent a single monomer, is in such a horizontal order always placed between the same two leaves (monomers) while there is exactly one internal node between any two adjacent leaves. This is exploited when an update is to be performed. Assume that the following update of the polymer configuration is proposed\footnote{In practice it is, of course, more efficient to modify the left part ($i=1,\dots,k-1$) if $k<N/2$. This is part of our implementation, but omitted here in favor of a simpler description.}
\begin{equation}
\mathbf{x}'_i=
        \begin{cases}
	\mathbf{x}_i & \text{if} \quad i\le k \\
	\mathcal{T}_U\bullet\mathbf{x}_i & \text{else}
        \end{cases}.
\end{equation}
One can then identify the internal node that is (in horizontal order) positioned between the leaves corresponding to $\mathbf{x}_k$ and $\mathbf{x}_{k+1}$ and use tree-rotations to move up this node until it becomes the root-node, i.e., the node on top without a parent. Since the horizontal order is preserved and the root node is between $\mathbf{x}_k$ and $\mathbf{x}_{k+1}$, it is clear that the leaves $\mathbf{x}_1,\dots,\mathbf{x}_k$  are now in the left part, i.e., the sub-tree to which the left child of the root-node is root, and the leaves $\mathbf{x}_{k+1},\dots,\mathbf{x}_N$ are in the right part. This situation is depicted in Fig.~\ref{fig:tree_upd}. Once the tree is in this shape, one can test whether the original left part overlaps with the transformed right part. If the spheres of the children of the root node do not overlap, which is the case if the distance between their midpoints is larger than the sum of the radii,
\begin{equation}
|\mathbf{y}_l-\mathcal{T}_U\bullet\mathbf{y}_r| > r_l+r_r,
\end{equation}
then there can be no overlap of any two individual monomers. Otherwise the resolution on one side has to be increased by stepping down one level in the tree. One has to either test that
\begin{eqnarray}
|\mathbf{y}_{ll}-\mathcal{T}_U\bullet\mathbf{y}_r| &>& r_{ll}+r_r \nonumber\\
{\rm and}\quad |\mathbf{y}_{lr}-\mathcal{T}_U\bullet\mathbf{y}_r| &>& r_{lr}+r_r
\end{eqnarray}
or that
\begin{eqnarray}
|\mathbf{y}_{l}-\mathcal{T}_U\bullet\mathbf{y}_{rl}| &>& r_{l}+r_{rl} \nonumber\\
{\rm and}\quad |\mathbf{y}_{l}-\mathcal{T}_U\bullet\mathbf{y}_{rr}| &>& r_{l}+r_{rr}.
\end{eqnarray}
Beforehand, either the transformation $\mathcal{T}_l$ or $\mathcal{T}_r$ has to be pushed down such that the actual positions $\mathbf{y}_{ll},\mathbf{y}_{lr}$ or $\mathbf{y}_{rl},\mathbf{y}_{rr}$ are used. This process is continued iteratively; whenever an individual inequality is violated, it has to be replaced by two conditions that are derived by splitting one of the participating nodes. Intuitively, one should split the larger one. For this model, it seems that splitting the node which contains more monomers is slightly more efficient ($\approx 1\%$) than splitting the node with the larger radius. The update is rejected if the process reaches a point where two nodes that are leaves, i.e., monomers, overlap. Otherwise, when the process terminates with all remaining inequalities fulfilled, the update is accepted and the transformation $\mathcal{T}_U$ is stored in the node $r$:
\begin{equation}
\mathcal{T}'_r=\mathcal{T}_U,
\end{equation}
or is multiplied to the existing transformation, if this node still contains one:
\begin{equation}
\mathcal{T}'_r=\mathcal{T}_U\circ\mathcal{T}_r.
\end{equation}
Finally, we retrace our steps and use the inverse tree-rotations as before, in order to rebalance the binary tree.

\begin{figure}
\begin{center}
\begin{tikzpicture}[level/.style={sibling distance=5em/#1,level distance=30pt}, every node/.style={transform shape},every node/.style={circle}]
\node [minimum size=0.6cm,draw, fill=green!20] (A) {B}
    child { node [minimum size=0.6cm,draw,fill=blue!20] {A} 
      child { node [minimum size=0.1cm] {} }
      child { node [minimum size=0.1cm] {} }
    }
    child { node [minimum size=0.6cm,draw,fill=green!20] {D} 
      child { node [minimum size=0.6cm,draw,fill=blue!20] {C} 
        child { node [minimum size=0.1cm] {} }
        child { node [minimum size=0.1cm] {} }
      }
      child { node [minimum size=0.6cm,draw,fill=blue!20] {E} 
        child { node [minimum size=0.1cm] {} }
        child { node [minimum size=0.1cm] {} }
      }
    };
\path[draw] (0,0.3) -- (0,0.7);
    
\node [right of=A,node distance=4cm,minimum size=0.6cm,draw, fill=green!20] (A) {D'}
    child { node [minimum size=0.6cm,draw,fill=green!20] {B'} 
      child { node [minimum size=0.6cm,draw,fill=blue!20] {A} 
        child { node [minimum size=0.1cm] {} }
        child { node [minimum size=0.1cm] {} }
      }
      child { node [minimum size=0.6cm,draw,fill=blue!20] {C} 
        child { node [minimum size=0.1cm] {} }
        child { node [minimum size=0.1cm] {} }
    }
      }
    child { node [minimum size=0.6cm,draw,fill=blue!20] {E} 
      child { node [minimum size=0.1cm] {} 
      }
      child { node [minimum size=0.1cm] {} 
      }
    };
\path[draw] (4,0.35) -- (4,0.7);

\draw[<->] (1.6,-1) -- (2.4,-1);

\end{tikzpicture}
\end{center}
\caption{\small{\label{fig:tree_rot} Tree-rotations are used to move nodes up or down. Only the nodes B and D are affected.}}
\end{figure}
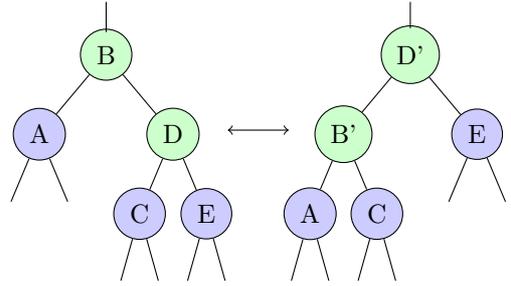

There are some differences to Clisby's technique beyond the mere transition from lattice to continuum. In particular, in our case the transformation in a given node applies to all nodes in the sub-tree to which it is root while in Ref.~\cite{Clisby1} transitions apply only to the sub-tree to which the right child of the node containing the transformation is root. This means that in our case for every monomer there are potentially $\log_2N$ transformations that need to be applied, while for the original version this number is smaller on average, e.g., there are no transformations that apply to the first monomer $\mathbf{x}_1$ in any case. On the other hand, since in our version we push down transformations, about half of  all nodes do not hold transformations at all such that the number that actually applies is smaller then the maximal value. In order to facilitate the tree-rotations we push down the transformations so that the relevant nodes are empty. This is not easily done in Clisby's version. Instead new transformations that keep the polymer configurations unchanged are determined which -- at least in one direction -- requires the inversion of one of the transformations. Since this is more complicated in continuum than on the lattice we chose this modification. On the other hand, with the original strategy it is in principle possible to omit the nodes containing individual monomers which would reduce the required memory by half \cite{Clisby4}. Achieving a similar improvement with our method would only be possible by using structurally different nodes that do not store transformations, radii, or sizes for leaves, which would render the code a bit more complicated. We have not compared both methods and do not claim that one performs better than the other.

\begin{figure}
\begin{center}
  \includegraphics[width=0.95\columnwidth]{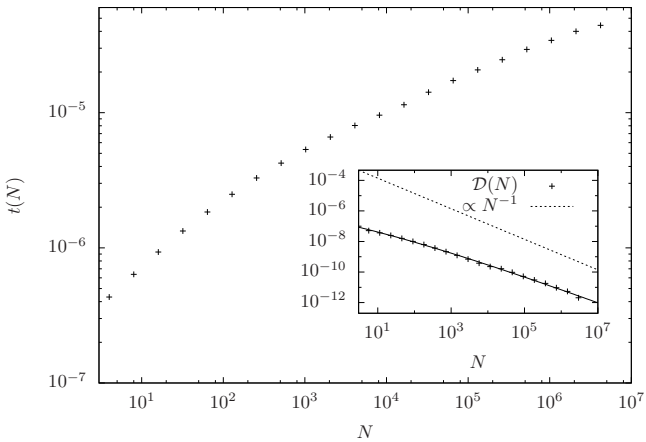}
\caption{\small{\label{fig:hs_times} The average time $t$ (in seconds) required for a single update for a hard-sphere polymer as a function of $N$. Inset: An estimate for the derivative $dt/dN\approx\mathcal{D}(N)=\left[t(N\sqrt{2})-t(N/\sqrt{2})\right]/\left(N/\sqrt{2}\right)$ with a fitted function as guide for the eye.}}
\end{center}
\end{figure}

Testing our implementation for $D=0.5$ with the pivot update we find that the desired efficiency is indeed achieved and that simulations with sizes up to $N\approx10^7$ are possible. In Fig.~\ref{fig:hs_times} we show the mean time $t(N)$ that is required for a single pivot update. Although we can not conclusively decide how this function behaves in the thermodynamic limit, it seems plausible that its derivative (inset of Fig.~\ref{fig:hs_times}) for large $N$ becomes proportional to $N^{-1}$ which would imply that $t(N)$ scales like $\log N$. Updates are accepted with satisfying probability which is, however, decreasing with system size. While $80\%$ of all updates are accepted for $N\approx 10$ this decreases to $25\%$ for $N=2^{22}$. We find that the rate of acceptance for pivot updates attempted at the center of the polymer is reasonably well described by $N^\kappa$ with $\kappa=-0.0926$.

\begin{figure}
\begin{center}
  \includegraphics[width=0.95\columnwidth]{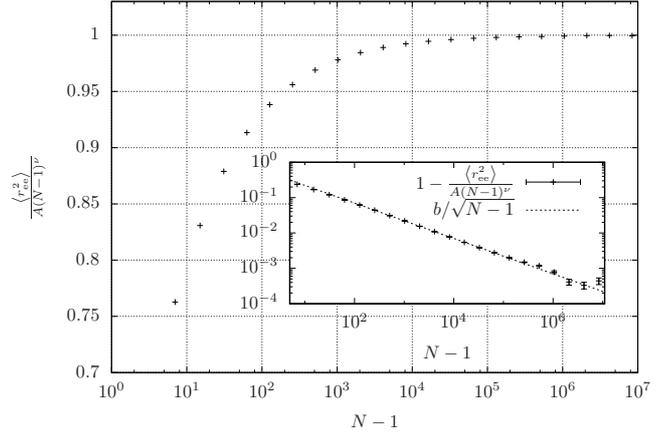}
\caption{\small{\label{fig:hs_ree} Scaling of the average squared end-to-end distance. Inset: Over several orders of magnitude in length $N-1$ the first correction to scaling is very well described by a single term with exponent $-0.5$.}}
\end{center}
\end{figure}

When looking at the data, we find that results agree with expectations. We acknowledge that analyzing the behavior of quantities like the squared end-to-end distance $\left\langle r_{\rm ee}^2 \right\rangle$ as a function of the number of bonds (or steps of a walk) $N-1$ is an intricate business and that the exponent of the first correction is not known. We might revisit it in detail in the future. With a brief glance at the case $D=1.0$ we notice that using the latest and most precise value for the Flory exponent available from a recent lattice study \cite{Clisby3}, $\nu=0.5875970(4)$, we find that
\begin{equation}
\left\langle r_{\rm ee}^2 \right\rangle=A(N-1)^{2\nu}\left(1-\frac b{\sqrt{N-1}}\right),
\end{equation}
with $A=1.77254$ and $b=0.701$ obtained from a fit for $N>10^3$, provides a decent description of the scaling (Fig.~\ref{fig:hs_ree}) with some room for improvement. This demonstrates very convincingly the expected universality of lattice and continuum self-avoiding walks.

For the notation used in this paper rotations are realized as matrices. However, we found that for sizes above $N=2^{16}$ at $D=1.0$ an implementation using quarternions shows a better performance.

\section{A parsimonious Metropolis algorithm}

Since the dawn of the information age the Metropolis algorithm \cite{Metropolis} has been the workhorse of computational statistical physics. Detailed balance,
\begin{equation}
P(\mu)W(\mu,\nu)=P(\nu)W(\nu,\mu),
\end{equation}
provides a solution to the Master equation. Here, $\mu,\nu$ are states of the system, $P$ occupation probabilities, and $W(\mu,\nu)$ is the probability to occupy $\nu$ in the next step if $\mu$ is occupied now. To be precise, the probability for a transition between two states in a Monte Carlo simulation is often the product of the probability of such an update being proposed and of it being accepted. Usually the probabilities for proposals are symmetric such that
\begin{equation}
P(\mu)W_{\rm accept}(\mu,\nu)=P(\nu)W_{\rm accept}(\nu,\mu),
\end{equation}
which in turn is solved by
\begin{equation}
W_{\rm accept}(\mu,\nu)=\min\left(1,\frac{P(\nu)}{P(\mu)}\right).
\end{equation}
If a canonical distribution at the inverse temperature $\beta=(k_{\rm B}T)^{-1}$ with $P(\mu)\propto \exp\left(-\beta E(\mu)\right)$ is aimed at, this becomes
\begin{equation}
W_{\rm accept}(\mu,\nu)=\min\left(1,e^{-\beta\Delta E}\right),
\end{equation}
with $\Delta E=E(\nu)-E(\mu)$. The usual procedure is to calculate $\Delta E$, determine this probability, and to draw a uniformly distributed random number $\xi\in[0,1)$. The update is accepted if $\xi<W_{\rm accept}(\mu,\nu)$. We intend to reverse this sequence. The last condition is equivalent to
\begin{equation}
\xi<e^{-\beta\Delta E}
\end{equation}
and consequently to
\begin{equation}
-\frac{\ln \xi}{\beta}>\Delta E,
\label{eq:Metr_inv_cond}
\end{equation}
assuming that $\beta>0$. Hence, it is possible to draw $\xi$ first and then to estimate $\Delta E$ with increasing precision until it can be decided whether (\ref{eq:Metr_inv_cond}) is fulfilled or violated. In cases where such an estimate can be done with less computational effort than a complete calculation, the simulation should run faster than with the standard technique, while with both methods the same updates are accepted or rejected and the trajectories through the configuration space are, therefore, identical.


\section{Simulating the Lennard-Jones polymer without cutoff}

For applying this idea to polymers with Lennard-Jones interaction the binary tree as used for the hard-sphere polymer again provides a well-suited data structure. Using trees in order to hierarchically estimate the interactions of $N$-body problems is not a new approach. Similar strategies have been used to simulate the (approximate) dynamics of gravitational systems as early as 1986 \cite{BarnesHut}. In these studies a more or less homogeneous three dimensional system is separated into cubic cells which are organized in an octree, where each internal node has eight children. The linear nature of the polymer simplifies the situation considerably. Again, distinct groups of monomers $A=\{\mathbf{x}_k,\dots,\mathbf{x}_{k+s_A-1}\}$ and $B=\{\mathbf{x}_l,\dots,\mathbf{x}_{l+s_B-1}\}$, with $k\ge 1,l\ge k+s_A,N\ge l+s_B-1$, are represented by spheres
\begin{eqnarray}
|\mathbf{x}_i-\mathbf{y}_A|&<&r_A,\quad i=k,\dots,k+s_A-1,\nonumber\\
|\mathbf{x}_j-\mathbf{y}_B|&<&r_B,\quad j=l,\dots,l+s_B-1.
\end{eqnarray}
The interaction between two such groups can be estimated if the distance between the spheres exceeds the minimum position of the Lennard-Jones potential
\begin{equation}
|\mathbf{y}_A-\mathbf{y}_B|-r_A-r_B>2^{1/6}\sigma.
\end{equation}
Due to the monotony of the interaction $U_{\rm LJ}(r)$ for $r>2^{1/6}\sigma$ the energy
\begin{equation}
E_{AB}=\sum_{i=k}^{k+s_A-1}\sum_{j=l}^{l+s_B-1} U_{\rm LJ}( |\mathbf{x}_i - \mathbf{x}_j| )
\end{equation}
has a lower bond
\begin{equation}
E_{AB}\ge s_A s_B U_{\rm LJ}(|\mathbf{y}_A-\mathbf{y}_B|-r_A-r_B)=E^{\rm min}_{AB}
\end{equation}
and an upper bound
\begin{equation}
E_{AB}\le s_A s_B U_{\rm LJ}(|\mathbf{y}_A-\mathbf{y}_B|+r_A+r_B)=E^{\rm max}_{AB}.
\end{equation}
The energy is minimal if all monomers are concentrated at the point closest to the opposite sphere and maximal if they are at the farthest point.\footnote{More sophisticated estimates are possible. For example, if one keeps track of the center of gravity of each group and if the distance between the spheres exceeds the inflection point of the Lennard-Jones potential, it can be shown that the energy is maximal if all monomers are concentrated in the centers of gravity of their respective groups. This energy is, therefore, also an upper bound. However, we found that the additional computations lead to a slower simulation in spite of the reduced depth resulting from the improved estimates.} If the estimate is not precise enough, it can be refined by splitting one of the contributing nodes $A\rightarrow \{A_\text{l},A_\text{r}\}$ or $B\rightarrow \{B_\text{l},B_\text{r}\}$ such that for instance
\begin{eqnarray}
E^{\rm min}_{AB}&=&E^{\rm min}_{A_\text{l}B }+E^{\rm min}_{A_\text{r}B}\quad \text{and}\nonumber\\
E^{\rm max}_{AB}&=&E^{\rm max}_{A_\text{l}B }+E^{\rm max}_{A_\text{r}B}
\end{eqnarray}
provides an improved estimate. This can be done in a recursive fashion similar to the process applied for the hard-sphere polymer. One of the interaction partners also has to be split up if the two spheres are too close to each other in order to allow for an estimation in the first place. For the Lennard-Jones polymer, single monomers are represented by spheres with zero radius, consequently the estimate becomes exact calculation,
\begin{equation}
E^{\rm min}_{AB}=E^{\rm max}_{AB}=U_{\rm LJ}( |\mathbf{x}_k - \mathbf{x}_k| ),
\end{equation}
if $s_A=s_B=1$, and it can be evaluated also for distances below the potential's minimum distance.

If the binary tree is prepared as was done previously (Fig.~\ref{fig:tree_upd}) with the root node possessing the children $l$ and $r$ and if we intend to modify the right part using the transformation $\mathcal{T}_{\rm U}$ such that symbolically $r\rightarrow r'=\mathcal{T}_{\rm U}r$ then
\begin{equation}
\Delta E\in [E^{\rm min}_{lr'}-E^{\rm max}_{lr},E^{\rm max}_{lr'}-E^{\rm min}_{lr}].
\end{equation}
Hence, the update is accepted if
\begin{equation}
E^{\rm max}_{lr'}-E^{\rm min}_{lr}<-\frac{\ln \xi}\beta
\label{eq:acc_cond_1}
\end{equation}
and rejected if
\begin{equation}
E^{\rm min}_{lr'}-E^{\rm max}_{lr}>-\frac{\ln \xi}\beta.
\label{eq:acc_cond_2}
\end{equation}
The interactions between $l$ and $r$ as well as $l$ and $r'=\mathcal{T}_Ur$ have to be evaluated and since in almost all cases initially the spheres will be too close or the estimates too rough, almost always interactions between nodes at lower levels will have to be included. Of course, we hope to avoid to consider too many interactions between small groups of monomers such that a decision is reached while interactions are evaluated at a low spatial resolution. This raises the question which particular interaction's estimate should be refined at any given point in order to improve the sum such that the overall process is efficient, i.e., terminates early. Similarly to the hard-sphere polymer it is possible to set up a recursive process that proceeds to smaller nodes until a particular condition is met. While previously non-intersection of the spheres was the only choice, now it is not so straightforward. Clearly the precision of an estimate of a node-node interaction $E_{AB}$ can be derived by just calculating the difference $E_{AB}^{\rm max}-E_{AB}^{\rm min}$, but since this scales with the product $s_As_B$ of the number of monomers of the two groups, it is useful to normalize
\begin{equation}
\alpha_{AB}\coloneqq \frac{E_{AB}^{\rm max}-E_{AB}^{\rm min}}{s_As_B},
\end{equation}
thus measuring the precision per monomer-monomer interaction. Now, we can define a target value $\alpha_c$ and descend to smaller nodes until only interactions that have smaller values $\alpha$ remain. If the result is not sufficiently precise for reaching a decision according to (\ref{eq:acc_cond_1}), (\ref{eq:acc_cond_2}) the process is repeated with a lower target value, e.g., $\alpha_c/2$. This approach has the advantage that it can easily be implemented using a recursive function and does not require additional data organization, since only the information about the particular interaction at hand is required.

An alternative, perhaps more intuitive and -- as it turns out -- more efficient strategy is to select the node-node interaction $E_{AB}$ that possesses the largest absolute uncertainty $E_{AB}^{\rm max}-E_{AB}^{\rm min}$ and split its larger node. However, since for large polymers there can be many millions of interactions, finding the most uncertain one is not entirely trivial. Note that the node-node interactions form two binary trees themselves. The roots are the interactions between the nodes $l,r$ and $l,r'$. Inner nodes in these trees represent estimates of interactions that at some point have been found to be too uncertain or impossible to make, due to close or intersecting spheres, and the current estimate of the total interaction energy is obtained by summation over the leaves. A tree is grown by adding two new interaction nodes to a former leaf, thus replacing its contribution in the total sum. These trees can be implemented as actual data structures and used as search trees in order to easily identify the leaf with the largest energy difference. This might be achieved by adding a link (pointer) to every node that points at the leaf with the largest energy difference in the sub-tree to which this node is root. Leafs point to themselves and whenever a leaf is modified, only the leaf itself and its direct ancestors have to be potentially modified by comparing the links in their children. If large polymers are considered with the number of monomers being in the range of several thousands, the trees can reach sizes of multiple millions of nodes. In order to limit the size of required memory we choose to limit the size of the tree and in those rare cases where the limiting size is reached we refrain from growing it further and proceed to improve the estimates in the leaves using the aforementioned recursive function and target precisions.

\begin{figure}
\begin{center}
  \includegraphics[width=0.95\columnwidth]{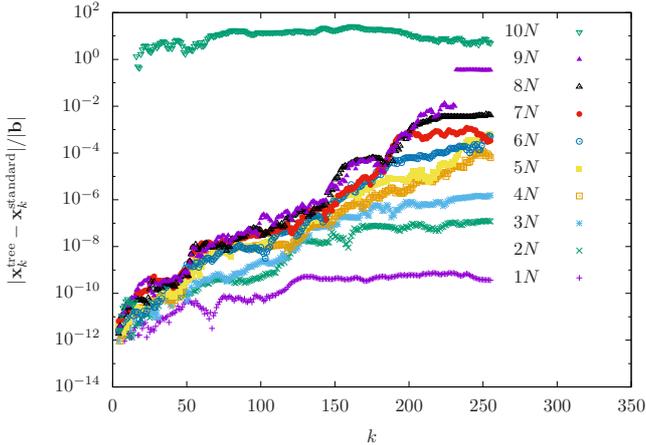}
\caption{\small{\label{fig:traj_comp} Difference in monomer positions for different methods (see text). Simplified algorithms that always modify the right part of the polymer were employed, hence the continued agreement for small $k$.}}
\end{center}
\end{figure}

Comparisons with a standard algorithm that calculates $\Delta E$ accessing monomer coordinates directly  have been done. Although our method should in principle create the same trajectory in state space, associativity as presented in (\ref{eqn:asso}) does not hold in a computer simulation. It makes a slight difference whether multiple transformations are sequentially applied to the monomer's coordinates or whether they are combined beforehand via the multiplication operation. The unavoidable rounding errors that occur in both cases lead to different results. Initially, these differences are tiny and do not affect whether an update is accepted or rejected, but after a short while the trajectories diverge. For an individual run for a polymer of size $N=256$ at $k_{\rm B}T=4\epsilon$, i.e., near the $\Theta$-point, using pivot and bond-rotation moves alternatingly, we found that it took about $10N$ such combined steps until differences in the monomer positions exceeded one bond length (Fig.~\ref{fig:traj_comp}). 

\begin{figure}
\begin{center}
  \includegraphics[width=0.95\columnwidth]{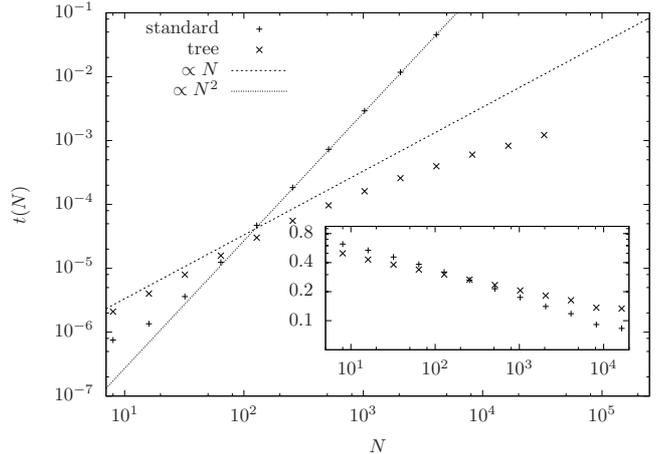}
\caption{\small{\label{fig:lj_times} Required times (in seconds) for the combination of a pivot and a bond-rotation update for the standard method and our algorithm at $k_{\rm B}T=4\epsilon$ for a Lennard-Jones polymer of length $N$  with untruncated interaction. Inset: Acceptance rate as function of $N$ for the pivot ($+$) and bond-rotation ($\times$) updates.}}
\end{center}
\end{figure}

Simulations for different sizes at the same temperature allows for comparisons and scaling of running times (Fig.~\ref{fig:lj_times}). It is of little surprise that the standard technique that requires the calculation of all modified monomer-mono\-mer distances soon approaches quadratic complexity. Our method is now substantially slower than for the hard-sphere polymer, but again it can be suspected that for large systems logarithmic scaling is realized. The acceptance rate is now more strongly affected by the polymer length and seems to decay in polynomial order with a larger albeit still favorably small exponent. We estimate $\propto N^{-0.28}$ for the pivot and $\propto N^{-0.18}$ for the bond-rotation update.

Once a function that estimates the interaction between two nodes in the tree has been set up, it can also be used to obtain an estimate for the energy of the entire polymer. If we define the energy of a node recursively as the sum of the interaction between its children and the energy of its children with individual monomers (leaves) possessing zero energy,
\begin{eqnarray}
\mathcal{H}_{k,\dots,k+s-1}&\coloneqq&\mathcal{H}_{k,\dots,k+\frac s2-1}+\mathcal{H}_{k+\frac s2,\dots,k+s-1}\nonumber\\
                             &&+\sum_{i=k}^{k+\frac s2-1}\sum_{j=k+\frac s2}^{k+s-1}U_{\rm LJ}(|\mathbf{x}_i-\mathbf{x}_j|),\nonumber\\
\mathcal{H}_{l}&\coloneqq&0,    
\end{eqnarray}
then the energy of the polymer is given by the energy of the root node
\begin{equation}
\mathcal{H}=\mathcal{H}_{1,\dots,N}.
\end{equation}
Since there are $N-1$ internal nodes, we need $N-1$ estimates of node-node interactions while each of which might recursively require multiple additional estimates. However, there is a good chance that for non-collapsed states and a target precision not too small this can be done with complexity $\mathcal{O}(N\log N)$ or faster. Once we can estimate the energy of one configuration the estimation of the difference between energies of two configurations is straightforward which allows to implement a replica-exchange algorithm \cite{parallel_temp1}. The exchange probability for swaps between two walkers at inverse temperatures $\beta_1>\beta_2$ is given by
\begin{equation}
P^{\rm accept}_{\rm swap}=\min\left(1,e^{(\beta_1-\beta_2)(E_1-E_2)} \right),
\end{equation}
so that the condition for accepting a replica-exchange update with reduced information reads
\begin{equation}
E_1-E_2>\frac{\ln\xi}{\beta_1-\beta_2}.
\end{equation}

\section{Simulating the Lennard-Jones polymer with cutoff}

Finally let us consider the case with a truncated potential which is the traditional technique to deal with large systems. The potential is set to zero beyond a certain cutoff-distance $r_c$ and the remaining part is shifted in order to avoid a discontinuity:
\begin{equation}
\tilde U_{\rm LJ}(r)=
        \begin{cases}
	U_{\rm LJ}(r)-U_{\rm LJ}(r_c) \quad \text{if} \quad r<r_c \\
	0 \quad \text{otherwise}
        \end{cases}.
\end{equation}

\begin{figure}
\begin{center}
  \includegraphics[width=0.95\columnwidth]{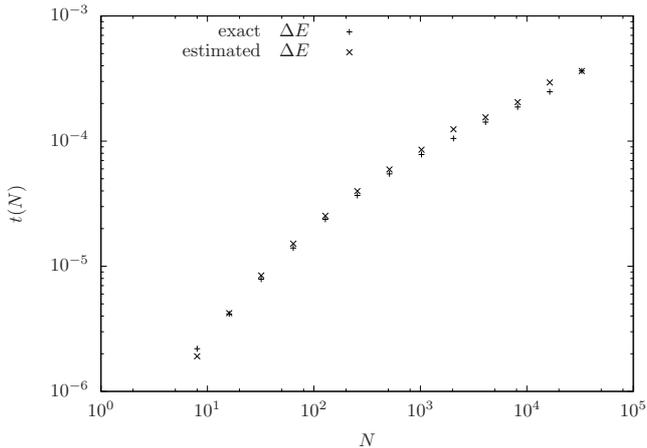}
\caption{\small{\label{fig:ljc_times} Time $t$ required for a combination of pivot and bond-rotation update as function of polymer size $N$ for the two versions of the Metropolis algorithm. The Lennard-Jones potential is truncated at $3\sigma$ and $k_{\rm B}T=3.5\epsilon$.}}
\end{center}
\end{figure}

There are two intuitive ways of simulating such a system using the binary-tree structure. During an update one could proceed similar to the hard-sphere case and establish which pairs of monomers are closer than the cutoff distance. Calculating their energies allows for a precise determination of $\Delta E$ and the standard Metropolis algorithm can be applied. Or the algorithm using reduced information could simply be used with the truncated potential. We would expect that the former is more efficient at low temperatures, since the estimation of interactions at small distances in dense configurations is less precise and many refinements might be necessary. At conditions near the collapse ($k_{\rm B}T=3.5\epsilon,r_c=3\sigma$), however, we find that both methods perform similarly well (Fig.~\ref{fig:ljc_times}). It is worth noting that in comparison to the untruncated case even for the largest systems considered ($N=2^{15}$) the introduction of the cutoff has only led to a threefold speedup.

\section{Conclusion and outlook}

In this study we have shown how the binary-tree method developed by Clisby for the simulation of self-avoiding walks on a lattice can adapted to hard-sphere polymers with continuous degrees of freedom. It turns out that system sizes of $N\approx10^7$ are not beyond the capabilities of these techniques. Although we can not be certain at this point, it seems that for very large systems the computational complexity of an individual Monte Carlo move scales like $\log N$.

We introduced a version of the Metropolis algorithm that does not rely on exact knowledge of the change in energy and reaches decisions based on sufficiently precise estimates. We applied this method to a Lennard-Jones polymer without any interaction-range cutoff close to the collapse transition and find that again a scaling of $\log N$ for single steps seems to be the asymptotic behavior and that polymers up to length $N\approx10^{4}$ can be investigated easily. The same idea in combination with the fact that estimates of the total energy can be obtained much faster than the exact value can be used to implement a replica-exchange algorithm for these systems.

Although an interaction-range cutoff of the potential as a means to enable a simulation in the first place is no longer required, polymers using such a truncated interaction can, of course, be simulated using these methods. We find that simulations are only modestly faster when a cutoff to the Lennard-Jones potential is used.

Since this work is intended to serve mainly as a proof-of-concept, relatively simple models were considered. However, it should be possible to introduce extensions like flexible spring-like bonds, bending stiffness or fixed bond angles, or multiple types of monomers with little effort.

\section*{Acknowledgements}
The project was funded by the Deutsche Forschungsgemeinschaft (DFG) through Collaborative Research Centre SFB/TRR 102 (project B04). We thank Nathan Clisby for helpful remarks.

\appendix
\counterwithin{figure}{section}

\section{Modifying central parts of the polymer}

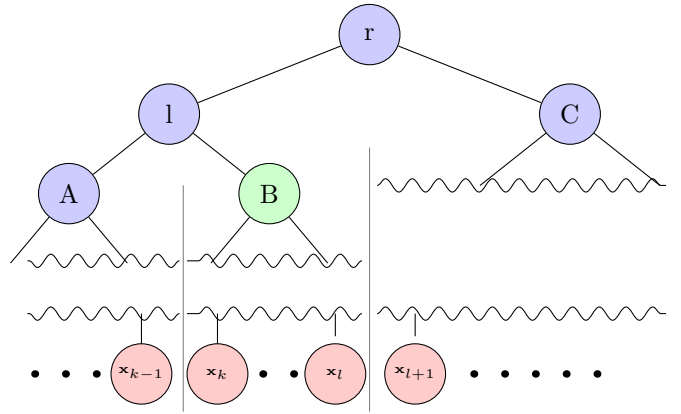
\begin{figure}
\begin{center}
\begin{tikzpicture}[level/.style={sibling distance=15em/#1,level distance=30pt}, every node/.style={transform shape},every node/.style={circle},circle dotted/.style={dash pattern=on .05mm off 4mm, line cap=round}]
\node [minimum size=0.8cm,draw, fill=blue!20] (A) {r}
    child { node [minimum size=0.8cm,draw,fill=blue!20] {l} 
      child { node [minimum size=0.8cm,draw,fill=blue!20] {A} 
        child { node [minimum size=0.1cm] {} }
        child { node [minimum size=0.1cm] {} }
      }
      child { node [minimum size=0.8cm,draw,fill=green!20] {B} 
        child { node [minimum size=0.1cm] {} }
        child { node [minimum size=0.1cm] {} }
      }
    }
    child { node [minimum size=0.8cm,draw,fill=blue!20] {C} 
        child { node [minimum size=0.1cm] {} }
        child { node [minimum size=0.1cm] {} }
    };
\path[draw,color=gray] (-2.45,-2.) -- (-2.45,-5);
\path[draw,color=gray] (0.,-1.5) -- (0.,-5);


\node [circle,minimum size=0.8cm,inner sep=0,draw,fill=red!20] at (-3.,-4.5) {${\scriptscriptstyle \mathbf{x}_{k-1}}$};
\path[draw] (-3.,-4.1) -- (-3.,-3.7);

\node [circle,minimum size=0.8cm,draw,fill=red!20] at (-2.,-4.5) {${\scriptscriptstyle \mathbf{x}_k}$};
\path[draw] (-2.,-4.1) -- (-2.,-3.7);

\node [circle,minimum size=0.8cm,draw,fill=red!20] at (-0.45,-4.5) {${\scriptscriptstyle \mathbf{x}_l}$};
\path[draw] (-0.45,-4.) -- (-0.45,-3.7);

\node [circle,minimum size=0.8cm,inner sep=0,draw,fill=red!20] at (0.6,-4.5) {${\scriptscriptstyle \mathbf{x}_{l+1}}$};
\path[draw] (0.6,-4.) -- (0.6,-3.7);


\draw[line width=1mm,circle dotted] (-4.4,-4.5) -- (-3.5,-4.5);
\draw[line width=1mm,circle dotted] (3.0,-4.5) -- (1.1,-4.5);
\draw[line width=1mm,circle dotted] (-1.4,-4.5) -- (-.8,-4.5);

\draw[snake it] (-0.1,-3.0)--(-2.4,-3.0);
\draw[snake it] (-0.1,-3.7)--(-2.4,-3.7);
\draw[snake it] (-2.5,-3.0)--(-4.5,-3.0);
\draw[snake it] (-2.5,-3.7)--(-4.5,-3.7);
\draw[snake it] (0.1,-2.)--(3.9,-2.);
\draw[snake it] (0.1,-3.7)--(3.9,-3.7);
\end{tikzpicture}
\end{center}
\caption{\small{ The binary tree during an update of the monomers $\mathbf{x}_k,\dots,\mathbf{x}_l$ which are at the highest level represented by node B. Node l and r limit this section on the left and right and node-node interactions (or overlaps) have to be considered between A,B and B,C.}}\label{fig:tree_upd_mid}
\end{figure}

As we have shown, for the Lennard-Jones polymer the acceptance rates decline when the system becomes large. It is, therefore, desirable to introduce additional updates of a more local nature that are not affected in this manner. This can be a crank-shaft move, where a part of the polymer is rotated around an axis passing through the limiting monomers or -- if a model with flexible bonds is used -- the shift of one or more adjacent monomers by a constant vector. The procedure is similar to the discussed update, with the distinction that now we have two internal nodes that pose the boundaries of the section that is to be moved. The tree is rearranged in a way that first one of them becomes root and in a second phase the other becomes a child of the new root node (Fig.~\ref{fig:tree_upd_mid}). With the tree in this shape the three relevant parts of the polymer are represented by single nodes and interaction between them can be evaluated recursively. Once the update is accepted it is again possible to apply the respective transformation at the highest level to a single node before rebalancing the tree.

\end{document}